\documentclass[12pt,a4paper]{article}
\usepackage{amsmath,amssymb,slashed}
\usepackage{amssymb,amsmath,bm}
\usepackage{color}
\usepackage{graphicx}

\newcommand{\as}{\alpha_s}

\newcommand{\MS}{\ensuremath{\overline{\text{MS}}}}

\long\def\symbolfootnote[#1]#2{\begingroup%
\def\thefootnote{\fnsymbol{footnote}}\footnote[#1]{#2}\endgroup}

\def \be{\begin{equation}}
\def \ee{\end{equation}}
\newcommand{\bea}{\begin{eqnarray}}
\newcommand{\eea}{\end{eqnarray}}
\def \nn{\nonumber}

\numberwithin{equation}{section}
\allowdisplaybreaks[4]

\begin{document}
\begin{titlepage}

\vspace*{3.5cm}

\begin{center}{\huge\bf\boldmath Revisiting  $B\to D \ell \nu$}
\end{center}
\vskip 2.5cm

\begin{center}
  {\bf Dante Bigi and Paolo Gambino, %\\[2mm]
   } \\[5mm]
  {\sl Universit\`a di Torino, Dip.\ di  Fisica \& INFN Torino, I-10125, Italy}
   \end{center}

\vskip 3cm

\begin{abstract}
We  re-examine the determination of $|V_{cb}|$ from $B\to D l \nu$ in view of recent experimental and theoretical progress, discussing 
the parameterization of the form factors and studying  the role played by 
the unitarity constraints. 
Our fit to experimental and lattice results for $B\to D \ell \nu$ gives  $|V_{cb}|=40.49(97)\ 10^{-3}$ and to $R(D)= 0.299(3)$. 
 \end{abstract}

\end{titlepage}

\section{Introduction}
The discrepancy between the determination of $|V_{cb}|$
from inclusive and exclusive semileptonic $B$ decays is
a long-standing problem in flavour physics. The CKM element $|V_{cb}|$
plays a central role in the analyses of CKM unitarity \cite{Bona:2006ah,Charles:2004jd}
and in the SM prediction of FCNC transitions, where its uncertainty is  often the dominant one \cite{Bobeth:2013uxa}. Its determination from inclusive decays 
is based on an Operator Product Expansion which parameterize the relevant non-perturbative physics in terms of non-perturbative constants  that are 
extracted from experiment, see \cite{Gambino:2015ima} for a review. A very recent analysis \cite{Gambino:2016jkc} points to
\be
|V_{cb}|= (42.00\pm 0.65)\ 10^{-3}.\label{incl}
\ee
The main channels for the exclusive determination of $|V_{cb}|$ have been so far $B\to D^{(*)}l \nu$, and until recently all analyses have focussed on the  zero-recoil point,
{\it i.e.}\ on maximal $q^2$.
 Indeed, in the heavy quark limit the relevant form factors are known exactly 
 at  zero-recoil  \cite{zr}, up to perturbative corrections,  and lattice calculations, which are anyway performed at high $q^2$, only need to determine the power-suppressed deviation from that limit. In  the $D^*$ case  the 
correction to the heavy quark  limit  is quadratically suppressed. The downside of zero-recoil analyses, however, is that  the decay rates vanish at zero-recoil (more rapidly in $B\to D l \nu$)
and that one therefore needs to extrapolate the experimental distributions, a problem which 
has been thoroughly addressed almost twenty years ago, with the introduction of
various model-independent parameterizations \cite{BGL1,BGL97,CLN}.

To date, the most precise exclusive determination of $|V_{cb}|$ is based on the calculation of the $D^*$ form factor at zero-recoil by the FNAL/MILC Collaboration \cite{Bailey:2014tva} and on the HFAG average \cite{hfag} of 
 B factory results analysed in the context of the CLN parameterization \cite{CLN}:
\be
|V_{cb}|= (39.04\pm 0.75)\ 10^{-3}.\label{excl}
\ee
This differs from (\ref{incl}) by 3$\sigma$, which becomes  2.8$\sigma$
once the QED corrections are treated in the same way in both cases.  It would be important to have  other independent lattice calculations, also in view of indications from
heavy quark sum rules that the form factor of  \cite{Bailey:2014tva} is overestimated \cite{Gambino:2012rd}.
Traditionally, the $D$ channel
has led to less precise determinations, mostly because of larger experimental errors, see for instance \cite{Bevan:2014iga}. 

The discrepancy between (\ref{incl}) and (\ref{excl}) is unwelcome. In principle,
it could signal new physics, as the $B\to D^*$  transition is sensitive only to the axial-vector component of the charged weak current. However,
this new physics would require new interactions which seem ruled out by electroweak constraints on the effective $Z\bar b b$ vertex \cite{Crivellin:2014zpa}. The situation is further complicated by the $3.9\sigma$ discrepancy \cite{hfag} between the  measurement of 
\be
R(D^{(*)})= \frac{{\cal B}(B\to D^{(*)}\tau \nu)}{{\cal B}(B\to D^{(*)}\mu \nu)},
\label{RD}
\ee
and their SM predictions, which depend on knowledge of the form factors in the whole available $q^2$ range. Different types of new physics could be responsible for this discrepancy, see \cite{Nandi:2016wlp} and Refs.\ therein for recent discussions.

In this context, any new information on $|V_{cb}|$ and on the semileptonic form factors is of great value. 
Two new elements have recently made the decays $B\to D \ell\nu$ more interesting in this respect. First, two  calculations of 
the form factors of  $B\to D \ell\nu$ {\it beyond} zero-recoil have appeared in 2015
\cite{Lattice:2015rga,Na:2015kha}. They represent the first unquenched calculations of these form 
factors performed at different $q^2$ values, which significantly reduces the uncertainty of the 
extrapolation from the $q^2$ region where most data are taken.
Second, a new, more precise  Belle measurement has been published \cite{Glattauer:2015teq}, which
for the first time provides the $q^2$ differential distribution with complete statistical and systematic  uncertainties and correlations. As we will show, the combination of these   
steps forward allows for a competitive extraction of $|V_{cb}|$
and for a very precise determination of the $B\to D$ form factors.

In this paper we revisit the  decay $B\to D \ell\nu$ in view of the above developments
and upgrade the unitarity bounds using recent three-loop calculations and up-to-date heavy quark masses.
In Section 2 we briefly review the model independent parameterization of the
form factors. We then update and discuss the impact of unitarity bounds, boosting them with 
the inclusion of higher states.  After presenting our inputs in Section 3, we perform a global fit of the available theoretical and 
experimental data, which leads to a precise determination of $|V_{cb}|$ and of the form factors. We also employ these form factors in the calculation of $R(D)$ and provide 
the most precise prediction to date. Section 5 summarizes our findings.

\section{Form factors and their parametrization}
The hadronic matrix element governing  the $B\to D \ell \nu$ decay is described by two form factors:
\be
\langle D(p')|V^{\mu}|\overline{B}(p)\rangle=f_{+}(q^2)(p+p')^{\mu}+f_{-}(q^2)(p-p')^{\mu} ,
\end{equation}
where $q^2=(p-p')^2$.
The differential rate can be written as:
\begin{equation}
\label{rate1}
\frac{d\Gamma}{dq^2}(B \rightarrow Dl{\nu_{l}})=
\frac{\eta_{ew}^2 G_{F}^2 \lvert {V_{cb}} \rvert ^2 m_B\lambda^{1/2}}{192 \pi^3 }\left(1-\frac{m_{l}^2}{q^2}\right)^2
 \Big[c_+^l\,
 %\frac{\lambda}{m_B^4} \left(1+\frac{m_{l}^2}{2q^2}\right)
 f_{+}(q^2)^2+
 %(1-r^2)^2 \frac{3m_{l}^2}{2q^2}
 c_0^l \,f_{0}(q^2)^2\Big]
\end{equation}
where $r=m_{D}/m_{B}$, $\lambda=(q^2-m_{B}^2-m_{D}^2)^2-4m_{B}^2m_{D}^2$,
\be
 %C=\frac{\eta_{ew}^2 G_{F}^2 \lvert {V_{cb}} \rvert ^2 m_B\lambda^{1/2}}{192 \pi^3 }\left(1-\frac{m_{l}^2}{q^2}\right)^2, 
c_+^l= \frac{\lambda}{m_B^4} \left(1+\frac{m_{l}^2}{2q^2}\right),
 \qquad
 c_0^l=(1-r^2)^2 \frac{3m_{l}^2}{2q^2},
 \ee
 and 
 \be
f_{0}(q^2)=f_{+}(q^2)+\frac{q^2}{m_{B}^2-m_{D}^2}f_{-}(q^2)\,, \nn
\ee
from which it follows that $f_+(0)=f_0(0)$.
In the limit of vanishing lepton mass the $f_{0}$ contribution becomes irrelevant. In fact,
it can be safely neglected except for decays into $\tau$ leptons. The factor $\eta_{ew}=1+\alpha/\pi \ln M_Z/m_B\simeq 1.0066$ \cite{sirlin}  takes into account the short distance QED corrections, namely  the electromagnetic running of the four-fermion operator from the weak to the $B$ scale and represents the leading electroweak correction. Unlike Ref.~\cite{Lattice:2015rga}  we do not include any Coulomb correction and expect the error due to subleading electroweak corrections to be negligible in comparison with other uncertainties.
The knowledge of $f_+$ and $f_0$ in the whole range $m_\mu^2\le q^2\le (m_B-m_D)^2$
allows for the calculation of $R(D)$, defined in (\ref{RD}).

The proper parameterization of the form factors $f_{+,0}(q^2)$ has been the subject 
of intense investigation, motivated in particular by  the need to extrapolate the information  
obtained in a restricted $q^2$ region to the whole $q^2$ range.  Lattice QCD 
calculations, for instance, are typically limited to the highest $q^2$ values. Here we consider the BGL, the  CLN, and the BCL parameterizations.

\subsection{The BGL parameterization}
The BGL parametrization was  originally proposed in \cite{BGL1} and further developed in   \cite{BGL97,BGL95_2}.
It follows from dispersion relations, analyticity, and crossing symmetry.
%without recourse to heavy quark simmetry.
In the case of semileptonic $B$ decays  $q^2$ ranges from $m_l^2$ to $(m_B-m_D)^2$
but the form factors can be continued analytically in the $q^2$ complex plane. They have 
a cut at $q^2=(m_B+m_D)^2$ and various poles corresponding to $B_c$ resonances with the appropriate quantum numbers.
Adopting the notation of \cite{BGL97}, we define
\begin{gather*}
t=q^2=(p-p')^2, \qquad t_{+}=(m_{B}+m_{D})^2,  \qquad t_{-}=(m_{B}-m_{D})^2,  \\
w=\frac{m_{B}^2+m_{D}^2-t}{2m_{B}m_{D}} \qquad z(w,{\cal N})=\frac{\sqrt{1+w}-\sqrt{2{\cal N}}}{\sqrt{1+w}+\sqrt{2{\cal N}}}, \qquad {\cal N}=\frac{t_{+}-t_{0}}{t_{+}-t_{-}},
\end{gather*}
where $z(t,t_0)$ maps the $q^2$ plane on a  unit disk.
The parameter $t_{0} < t_+$ is a free parameter which determines the point in the $q^2$-plane to be mapped  onto the origin of the $z$  plane by the conformal transformation $q^2\to z$. 
The two form factors $f_+$ and $f_0$  are parameterized by:
\bea
\label{fp}
f_{+}(z)&=&\frac{1}{P_{+}(z) \phi_{+}(z,{\cal N})}\sum_{n=0}^{\infty}a_nz^{n}(w,{\cal N})\\
\label{f0}
f_{0}(z)&=&\frac{1}{P_{0}(z) \phi_{0}(z,{\cal N})}\sum_{n=0}^{\infty}b_nz^{n}(w,{\cal N}),
\eea
where $P_{+,0}(z) $ are known as {\it Blaschke factors} and $\phi_{+,0}(z)$ as {\it outer functions}. They will be introduced shortly. In practice one truncates the series in (\ref{fp},\ref{f0}) at some maximal power $N$. In our fits we will consider $N=2,3,4$.

The main advantage of the BGL class of parameterizations is that the parameters $a_{n}$ of equation  (\ref{fp})  are constrained by unitarity conditions
\be
\sum_{n=0}^\infty a_n^2<1,\qquad \qquad\qquad\sum_{n=0}^\infty b_n^2<1,
\label{weakunit}
\ee
 that follow from analyticity, 
crossing symmetry,  and quark-hadron duality, and  will 
be discussed in detail in the next Section.  Indeed, it is possible to write dispersion relations
for the correlator of two  flavour changing currents evaluated at $q^2=0$, where it can be computed reliably in perturbation theory because the heavy quark masses are much larger than the QCD scale.
Assuming global quark-hadron duality, the dispersion relations relates integrals of the 
form factors at $q^2$ values outside the physical range to the perturbative calculation
of the correlator. The unitarity bounds of (\ref{weakunit}) then follow,  see \cite{Boyd:1997qw} for a pedagogical introduction.

In the case of $B\to D$ semileptonic decays with massless leptons $z$ can vary   between
\begin{equation*}
z_{min}=-\frac{\sqrt{\cal N}-1}{\sqrt{{\cal N}}+1} \qquad {\rm and} \qquad z_{max}=\frac{1+r-2\sqrt{{\cal N}r}}{1+r+2\sqrt{{\cal N}r}}.
\end{equation*}
Choosing  $t_{0}=t_{-}$, one has ${\cal N}=1$; the range of variation of $z$ is  $ 0 \le z \le \frac{(1-\sqrt{r})^2}{(1+\sqrt{r})^2} \simeq 0.0646$ and the point of zero recoil ($w=1$) is at $z=0$.
This is the most common choice in the literature.
Another convenient choice for $t_0$ is the one which leads to a symmetric $z$-range
using the  condition $|z_{min}|=|z_{max}|$. This corresponds to  $t_{0}=t_{+}-{\cal N}(t_{+}-t_{-})$ and ${\cal N}=\frac{1+r}{2\sqrt{r}}$.  With this prescription the 
maximum physical value of $|z|$ is minimized and  $z_{max}\simeq 0.032$. 
While in principle a smaller range in $z$, combined with (\ref{weakunit}), forces a 
faster convergence of the $z$-expansion, we have checked that in our case this choice brings  no advantage with respect to setting $t_0=t_-$. This is likely due to the 
precise lattice QCD constraints we employ.
From now on we will only consider the case $t_0=t_-$ and $z$ will always stand for $z(w,1)$.

 In our case only the transverse and longitudinal parts of the vector current correlator are relevant
\be
\left(-g^{\mu\nu}+\frac{q^\mu q^\nu}{q^2}\right) \Pi^T(q^2) +\frac{q^\mu q^\nu}{q^2}\Pi^L(q^2)\equiv
i\int d^4x \,e^{i qx}\langle 0|TJ^\mu(x) J^{\dagger\nu}(0)|0\rangle
\ee
with $J^\mu=\bar{c}\gamma^\mu b$. The longitudinal and transverse part correspond
to spin 0 and spin 1, respectively. The derivatives
\be
\chi^L(q^2)=\frac{\partial \Pi^L}{\partial q^2}, \qquad 
\chi^T(q^2)=\frac12 \frac{\partial^2 \Pi^T}{\partial (q^2)^2},\nn
\ee
 satisfy unsubtracted dispersion relations on which the unitarity bounds are built. The value $q^2=0$ is sufficiently far from the threshold region and is generally employed. 

For the perturbative evaluation of $\chi^{L,T}(0)$ we update Ref.~\cite{BGL97} using recent values of the heavy quark masses and the $O(\as^2)$ corrections computed in \cite{Grigo:2012ji}. We neglect all contributions of condensates, which have been shown to be negligible.
Several precise determinations of the bottom and charm masses have appeared in recent 
years, see \cite{pdg} for a review. Here we first use the $\MS$ values $\overline{ m}_b(\overline{ m}_b)=4.163(16)$GeV, $\overline{ m}_c(3 {\rm GeV})=0.986(13)$GeV
from \cite{Chetyrkin:2009fv}
and $\alpha_s^{(5)}(\overline{ m}_b(\overline{ m}_b))=0.2268(23)$ and get
\bea
&&\chi^{T}(0)=[5.883+0.552_{\as}+0.050_{\as^2}] \ 10^{-4}\,{\rm GeV}^{-2}=6.486(48)\  10^{-4}\,{\rm GeV}^{-2} , \nn \\
&& \chi^{L}(0)=[5.456+0.782_{\as}-0.034_{\as^2}]  \ 10^{-3}=6.204(81) \ 10^{-3}
\label{chi0MS}
\eea
where the errors reflect only the uncertainties on the input quark masses and we have neglected the small correlation between them. The effect of the $O(\as^2)$ corrections is less than 1\%, but the values differ significantly 
from those used  in \cite{BGL97}, $\chi^T(0)=4.42 \ 10^{-4}\,{\rm GeV}^{-2}$ and $\chi^L(0)=4.07 \ 10^{-3}$,  because of the different, more precise inputs.
An alternative determination is obtained using the fit to semileptonic moments of \cite{Alberti:2014yda} and employs the kinetic $b$ mass $m_b^{kin}(1{\rm GeV})=4.553(20)$GeV, $\overline{ m}_c(3 {\rm GeV})=0.987(13)$GeV and $\alpha_s^{(5)}( m_b^{kin}(1{\rm GeV}))=0.2208(22)$, leading to 
\bea
&&\chi^{T}(0)=[4.958+1.059_{\as}+0.309_{\as^2}]\  10^{-4}\,{\rm GeV}^{-2}=6.326(51)\  10^{-4}\,{\rm GeV}^{-2}, \nn \\
&& \chi^{L}(0)=[5.905+0.564_{\as}-0.136_{\as^2}]\ 10^{-3}=6.332(74) \ 10^{-3},\label{kin}
\eea
where we have taken into account the correlation between $m_b$ and $m_c$ from the fit 
of \cite{Alberti:2014yda}.  The NNLO corrections amount to $+5.1$ and $-2.1$\%, respectively.
Because of the off-shell nature of $\chi^{T,L}(0)$,  the $O(\as)$ and $O(\as^2)$ corrections are more sizeable when we employ the kinetic mass for the $b$ quark instead of the $\MS$ mass; taking into account the theoretical uncertainty due to higher order corrections, larger 
in the second case,   Eqs.~(\ref{chi0MS}, \ref{kin}) are perfectly consistent. 
In the following, we adopt the more precise values in (\ref{chi0MS}) as our reference. 
\begin{table}
\begin{center}
\begin{tabular}{c|c|c|c} 
\hline
Type & Mass (GeV) & Decay constants (GeV) & Ref. \\
\hline
$1^-$ & 6.329(3) & 0.422(13) & \cite{pdg,Dowdall2012,Colqu2015}  \\
\hline
 $1^-$ & 6.920(20)  & 0.300(30)  &  \cite{pdg,Dowdall2012,Ikh2005}   \\
\hline
$1^-$ & 7.020 &  & \cite{Rai2013} \\
\hline
$1^-$ & 7.280 & &  \cite{Eichten}  \\
\hline
$0^+$ & 6.716  & & \cite{Rai2013} \\
\hline
$0^+$ & 7.121 & & \cite{Rai2013}  \\
\hline
\end{tabular}
\end{center}
\caption{Relevant $B_c$ masses and decay constants.}
\label{Pole Mass Values}
\end{table}

The $1^-$ $B_c$ resonances below the $BD$ pair production threshold contribute as single particles to the unitarity sum. Their effect can be effectively seen as a reduction of $\chi^{T,L}(0)$, see 
\cite{CLN}: 
\begin{equation}
\tilde\chi^{T}(0) = \chi^{T}(0)-\sum_{n=1,2}\frac{f_{n}^2(B_{c}^*)}{M_{n}^4(B_{c}^*)}
\end{equation}
where $f_{n}$ are the decay constants and $M_{n}$ the masses of the $B_{c}^*$ mesons.
The decay constant is strongly suppressed for $0^+$ states and we therefore neglect this contribution to the scalar channel. We likewise do not consider poles above the $BD$ threshold to avoid double-counting.
In Table \ref{Pole Mass Values}  the relevant $B_c$ masses and decay constants are presented with their sources, among which are  recent lattice calculations.
The masses will also be  used to evaluate the Blaschke factors later on.
A conservative  10\%  uncertainty is assigned to $f_{2}(B_{c}^*)$. Since the unitarity bounds
emerge from the assumption that a single channel, or a set of channels, saturate the 
dispersion relation,  larger values of $\chi^{T,L}$ constrain the form factors less effectively.
Therefore, to be  conservative we compute $\tilde \chi^{T}$ using the decay constants reduced by one standard deviation 
and the mass values increased by one standard deviation. The result in units of GeV$^{-2}$ is
\begin{equation}
\tilde\chi^{T}(0) =6.486 \times 10^{-4}-\frac{(0.409)^2}{(6.332)^4}-\frac{(0.270)^2}{(6.940)^4}=5.131 \times 10^{-4}\label{chitilde}
\end{equation}

Another ingredient in Eqs.~(\ref{fp},\ref{f0}) are the Blaschke factors
\be
\label{p+0}
P_{+}(z)=\prod_{P_+=1}^{3}\frac{z-z_{P_+}}{1-zz_{P_+}}, \qquad \quad
P_{0}(z)=\prod_{P_0=1}^{2}\frac{z-z_{P_0}}{1-zz_{P_0}}   
\ee
where  $z_{P}$ is defined as:
\begin{equation*}
z_{P}=\frac{\sqrt{t_{+}-m_{P}^2}-\sqrt{t_{+}-t_{0}}}{\sqrt{t_{+}-m_{P}^2}+\sqrt{t_{+}-t_{0}}},
\end{equation*}
where $m_P$ represents the location of a $B_c$ narrow resonance.
The product is extended to all the $B_c$ resonances below the $B$-$D$ threshold with the appropriate quantum numbers ($1^-$ for $P_+$ and $0^+$ for $P_0$).
The Blaschke factors remove the subthreshold poles from the form factors, making the form factors analytic for all $q^2$ values below the cut. 

Finally, the outer functions reflect the way in which 
the form factors enter the dispersive integral, which depends on the helicity amplitude they belong. Their normalization depends on $\chi^{T,L}$ because we want to have the unitarity bounds in the simplest form (\ref{weakunit}). 
The  outer functions $\phi_{+,0}$ are given by 
\bea
\label{fi+}
&& \hspace{-1.3cm}\phi_{+}(z)=%10.901
k_+
 \ \frac{(1+z)^2\sqrt{1-z}}{[(1+r)(1-z)+2 \, \sqrt{r} \, (1+z)]^5} ,
 \qquad k_+=\frac{8\,r^2}{m_B} \sqrt{\frac{8n_I}{3\pi \tilde\chi^T(0)}}\simeq 12.43,
 \\
\label{fi0}
&& \hspace{-1.3cm}\phi_{0}(z)=%8.866 
k_0 \ \frac{(1\!-\!z^2)\sqrt{1-z}}{[(1+r)(1-z)+2 \, \sqrt{r} \, (1+z)]^4}, \qquad \
k_0=  r(1\!-\!r^2)\sqrt{\frac{8n_I}{\pi \chi^L(0)}}\simeq 10.11,
\eea
where %$z=z(w,1)$. 
$n_I$ is a factor  
that %, for the $B\to D^{(*)}$ transitions, 
simply counts the number of massless spectator quarks. 
Although SU(3) breaking appears to be  small in form factors calculations,  we prefer to be conservative and use $n_I=2.6$. 
Replacing   (\ref{p+0}-\ref{fi0}) 
%or  (\ref{p+2})-(\ref{fi02})  
in  equations  (\ref{fp},\ref{f0}) we obtain the BGL  parameterizations of $f_{+}$ and $f_{0}$.
There appears to be
some confusion on $k_0$ in the literature (see {\it e.g.}\ \cite{Glattauer:2015teq,Lattice:2015rga}),
possibly generated by the unusual definition of $f_0$ in \cite{BGL97}. We stress  that the
precise definition and inclusion of both $P_{+,0}$ and $\phi_{+,0}$ is instrumental to using 
the unitarity bound on the sum of the squared coefficients of the $z$-expansion. Without
this bound there is no difference between the $z$-expansion and any other power
expansion of the form factors.

\subsection{CLN parameterization}

The CLN  parametrization was   proposed in \cite{CLN}  and has been  extensively  used in the literature. % to fit experimental and lattice results.
It is also based on dispersion relations and unitarity but  it additionally exploits
Heavy Quark Effective Theory (HQET) to reinforce the unitarity bounds.
Indeed, the form factors of the two-meson states contributing to the two point function are related by heavy quark symmetry and in the heavy quark limit either vanish or are proportional to the Isgur-Wise function \cite{BGL97}.   Ref.~\cite{CLN} also includes ${\cal O}(1/m)$ heavy quark symmetry breaking corrections, computed with input from light-cone sum-rules, and leading short distance corrections to these relations. We will describe in more detail the method in Section 2.4.

The reinforced unitarity bounds allow Ref.~\cite{CLN} to establish approximate relations between the slope and the higher power coefficients of the reference  form factor $f_+$, and to provide simplified formulas
valid within $\approx 2\%$. For instance, our form factors of interest
are expressed in terms of two parameters only,
\bea
\label{fpCLN}
f_{+}(z)&\simeq& %\frac{  1+r}{2 \sqrt{r}}  \,
f_+({0})\left[1-8  \rho^2_{1}  z+(51  \rho^2_{1}-10)  z^2-(252  \rho^2_{1}-84)  z^3)\right]\\
\label{f0CLN}
\frac{f_{0}(z)}{f_{+}(z)}&\simeq &\left(\frac{2\sqrt{r}}{1+r}\right)^2  \frac{1+w}2   
 \,1.0036 \left[1-0.0068  w_1 +0.0017 w_1^2 -0.0013 w_1^3 \right]
\eea
where $w_1=w-1,$. Notice in particular that the  ratio $f_0/f_+$ is fixed by the NLO HQET calculation implemented in \cite{CLN}. All other $B^{(*)}\to D^{(*)}$ form factors are 
similarly expressed as $f_+(z)$ times a ratio computed at NLO in HQET.
One should bear in mind that the heavy mass expansion here is an expansion in $1/m_c$ and therefore all form factors in this parameterization are subject, in principle, to ${\cal O}(1/m_c^2)\sim 5-10\%$ corrections. Indeed, the ratio of the $B\to D^*$ and $B\to D$ form factors at zero recoil is 0.948 at NLO in HQET \cite{CLN}, while the most precise
lattice calculations lead to 0.860(14) \cite{Bailey:2014tva,Lattice:2015rga}.
However, as long as the CLN parameterization is used to describe the shape of a  single form factor, like in (\ref{fpCLN}), it provides a simple and effective parameterization,  unless of course the experimental or theoretical constraints reach the $\sim 1\%$ precision. 

\subsection{The BCL parameterization}
The  BCL parameterization   \cite{Bourelly2008} was proposed to overcome problems 
that appear due to the truncation of the BGL expansion.  When the BGL expansion is truncated at some finite power $N$, 
the form factor develops an unphysical  singularity at the threshold $t_+$ and behaves at large $|q^2|$ in contradiction with perturbative QCD scaling.
While these problems are related to the behaviour of the form factor at values of $q^2$ much larger than those relevant for $B\to D\ell\nu$ and are therefore likely to be irrelevant in the present context, the BCL parameterization offers a simple
alternative to BGL that avoids these potential shortcomings. The 
two form factor of interest are given by 
\bea
f_+(q^2)&=&\frac1{1-q^2/M_+^2} \sum_{k=0}^{N} a_k \left[z^k -(-1)^{k-N-1}\frac{k}{N+1} z^{N+1}\right],\label{BCL1}
\\
f_0(q^2)&=&\frac1{1-q^2/M_0^2} \sum_{k=0}^{N} b_k z^k,\label{BCL2}
\eea
where $M_{+,0}$ are the masses of the two closest $B_c$ resonances in the $+,0$ channels, see Table 1. The $z^{N+1}$ terms in (\ref{BCL1}) ensure the proper behaviour of the form factor at $q^2\approx t_+$. There is no point in introducing additional pole terms for
resonances that lie even further. The unitarity bounds we have considered above 
can be mapped onto the BCL parameters as shown in \cite{Bourelly2008}. As they assume a more complicated form, we will only consider the weak bounds for the BCL parameters.

\subsection{Strong unitarity constraints}
The unitarity bounds of Eq.~(\ref{weakunit}) assume that a single hadronic channel, in our case $BD$,  saturates
the bound; we can label them  {\it weak unitarity bounds}. 
Of course there are a number of additional two body channels ($BD^*,B^*D,B^*D^*, \Lambda_b \Lambda_c,\dots$) with the right quantum numbers, as well as higher multiplicity  channels, that give  positive contributions to the absorptive part 
to the two-point function and can  strengthen the unitarity bound on the coefficients of the $BD$ form factors.

In the case of the four states $\bar{B}^{(*)}\bar{D}^{(*)}$, one can use heavy quark symmetry to connect   the form factors of the various channels.
The implementation of these relations in the unitarity conditions  is outlined in \cite{BGL97}:
only amplitudes of  fixed spin parity enter each dispersion relation, leading to the 
{\it strong}  unitarity condition:
\begin{equation}
\label{sumstrong}
\sum_{i=0}^{H}\sum_{n=0}^{\infty}b_{in}^2 \le 1.
\end{equation}
Here all  helicity amplitudes $i=0 \dots H$ for processes involving $\bar{B}^{(*)}\bar{D}^{(*)}$  with the right quantum numbers  must be  included.

We follow here  the approach proposed in \cite{BGL97} to derive strong unitarity bounds on
the coefficients of $f_+$, but we include short distance and $1/m$ corrections to the heavy quark limit as done in \cite{CLN}. One can compute strong constraints for the coefficients
of $f_0$ as well, but they would play a marginal role in our analysis. We will therefore
use  only the weak unitarity constraint for the coefficients of the scalar form factor.

Considering the vector ($1^{-}$) form factors for ${B}^{(*)}{D}^{(*)}$ states, 
there are seven helicity amplitudes, $H=7$. Each form factor $F_{i}$ can be put in the general form:
\begin{equation}
F_{i}=\tilde{f_{i}}\sum_{n}b_{in}z^{n}
\end{equation}
where $\tilde{f_{i}}(z)=1/P_{i}(z) \phi_{i}(z)$ are  known functions.
%It follows  that:
%\begin{equation}
%%\sum_{n}b_{in}z^{n}=\frac{F_{i}}{\tilde{f_{i}}} \qquad 
%%\sum_{i=1}^{7} 
%\sum_{n}b_{in}z^{n}=%\sum_{i=1}^{7}
%\frac{F_{i}}{\tilde{f_{i}}}
%\end{equation}
If $b_{1n}=a_n$  are the  coefficients of $f_{+}(z)$, it is possible to   rearrange the  $z$ expansion of each form factor as
\begin{equation}
\label{strun}
 \sum_{n}b_{in}z^{n}=\sum_{n}a_{n}z^{n}
c_i(z)
%c_{i1}(z)\frac{\tilde{f_{+}}}{\tilde{f_{i}}}
\end{equation}
where $c_{i}(z)=F_{i} \tilde{f}_+/(f_{+} \tilde{f}_i)$ 
%$c_{i1}(z)=\frac{F_{i}}{f_{+}}$ 
and  only the $f_{+}$ coefficients appear explicitly.
In order to  use equation  (\ref{strun}) to obtain a unitarity condition involving only the $a_{n}$, an explicit approximation for the  $c_{i}(z)$  is required, {\it i.e.} 
information on the other form factors is needed. Once the $c_i(z)$ are known, their $z$-expansion  allows for the determination of the coefficients $b_{in}$ to be employed in
(\ref{sumstrong}). Of course, the maximum reached by the index $n$  in (\ref{sumstrong})
coincides with the maximum power  of $z$ we employ in (\ref{fp},\ref{f0}), namely $N$.

In Ref.~\cite{BGL97}  the exact heavy quark limit in used in order to fix the functions $c_{i}(z)$. 
%the parameter N  is   employed  to  optimize the constraints and  equation (\ref{bnan}) is enforced up to the second derivative. Here an alternative  approach is proposed:
Here we evaluate them incorporating the $1/m$ heavy quark symmetry breaking and
short distance corrections as done in \cite{CLN}.
%This  allows to include  an evaluation  of the  $1/m$ corrections  to  the exact heavy quark spin simmetry used in  the original reference;
% The $c_{i}(z)$ functions are evaluated 
%\item  The N parameter is assumed equal to 1 and the  power series are developed in z=0;
In the notation of \cite{BGL97} the form factors  involved
are $F_i=(f_{+}, g, \hat{g}, V_{++}, V_{+0}, V_{0+}, V_{00})$.  They are related to those in
\cite{CLN} by 
%In order to implement the first step,  the form factors defined in \cite{BGL97} and have to be  related.    Inspecting the definitions  and assuming the same  notation of the two references the following relations are derived:
\begin{gather*}
\label{BoydCapri}
f_{+}=\frac{m_{B}+m_{D}}{2\sqrt{m_{B}m_{D}}}V_{1}, \quad  g=\frac{V_{4}}{\sqrt{m_{B}m_{D^{*}}}} ,\quad  \hat{g}=-\frac{V_{5}}{\sqrt{m_{B^{*}}m_{D}}}, \\
V_{+0}=\frac{V_{6}}{\sqrt{m_{B^{*}}m_{D^{*}}}} ,\quad V_{++}=\frac{V_{7}}{\sqrt{m_{B^{*}}m_{D^{*}}}}, \quad V_{0+}=-\frac{m_{B^{*}}+m_{D^{*}}}{\sqrt{2m_{B^{*}}m_{D^{*}}}} V_{2} ,\quad V_{00}=\frac{m_{B^{*}}+m_{D^{*}}}{2\sqrt{m_{B^{*}}m_{D^{*}}}} V_{3}.
\end{gather*}
\begin{table}[t]
\begin{center}
\begin{tabular}{c|c}\hline
Mass & Value (GeV) \\
\hline
$m_{B}$ & 5.27942 \\
\hline
$m_{D}$ & 1.86723 \\
\hline
$m_{\mu}$ & 0.1057 \\
\hline
$m_{\tau}$ & 1.7768 \\
\hline
$m_{B^{*}}$ & 5.325\\
\hline
$m_{D^{*}}$ & 2.009\\
\hline
\end{tabular}
\caption{Mass values employed in the paper.}
\label{tab:masses}
\end{center}
\end{table}
Using  these relations and the ratios $V_i/V_1$ computed  in \cite{CLN} at NLO in HQET and given in their Appendix,
it is easy to find the relevant $c_i(z)$.
%
%for the $c_{i1}(z)$ functions the following expressions:
%\bea
%\label{ci1Capri}
%&&c_{11}=1 \nn\\ &&c_{21}=1.2179\left(1-0.1428 w_1-0.0015w_1^2+0.0004w_1^3\right)\frac{2\sqrt{m_{D}}}{\sqrt{m_{D^{*}}}(m_{B}+m_{D})}\nn \\
%&&c_{31}=-1.0676\left(1-0.0362 w_1-0.0015 w_1^2+0.0004 w_1^3\right)\frac{2\sqrt{m_{B}}}{\sqrt{m_{B^{*}}}(m_{B}+m_{D})}\nn \\
%&&c_{41}=1.3416\left(1-0.1987w_1-0.0015w_1^2+0.0004w_1^3\right)\frac{2\sqrt{m_{B}m_{D}}}{\sqrt{m_{B^{*}}m_{D^{*}}}(m_{B}+m_{D})} \nn\\
%&&c_{51}=1.4919\left(1-0.2278w_1-0.0015w_1^2+0.0004w_1^3\right)\frac{2\sqrt{m_{B}m_{D}}}{\sqrt{m_{B^{*}}m_{D^{*}}}(m_{B}+m_{D})} \nn\\
%&&c_{61}=-1.0681\left(1-0.1944w_1\right)\frac{\sqrt{2}(m_{B^{*}}+m_{D^{*}})\sqrt{m_{B}m_{D}}}{\sqrt{m_{B^{*}}m_{D^{*}}}(m_{B}+m_{D})} \nn\\
%&&c_{71}=1.1361\left(1-0.2474w_1\right)\frac{(m_{B^{*}}+m_{D^{*}})\sqrt{m_{B}m_{D}}}{\sqrt{m_{B^{*}}m_{D^{*}}}(m_{B}+m_{D})} 
%\eea
Indeed, the ratios $\tilde{f}_+/{\tilde{f}_{i}}$ are easily computed from  \cite{BGL97}.   
Depending on the specific form factor either 3 or 4 poles have to be subtracted with the Blaschke factors because of the different $t_{+}$ thresholds.

The strong unitarity bounds are finally obtained using Tables \ref{Pole Mass Values} and \ref{tab:masses}, and Eq.~(\ref{chitilde}):
\begin{itemize}
\item $N=2\\
442.82 a_0^2 - 101.619 a_0 a_1 + 
 34.947 a_1^2 - 127.668 a_0 a_2 + 
 33.234 a_1 a_2 + 16.4754 a_2^2\le 1
%453.195 a_{0}^2 - 99.9664 a_{0} a_{1} + 34.7346 a_{1}^2 - 
% 130.601 a_{0} a_{2} + 33.0545 a_{1} a_{2} + 16.9168 a_{2}^2 
 $
\item $N=3 \\
1707.54 a_{0}^2 + 1299.57 a_{0} a_{1} + 442.82 a_{1}^2 - 
 356.01 a_{0} a_{2} - 101.62 a_{1} a_{2} + 34.947 a_{2}^2 - 
 206.767 a_{0} a_{3} - 127.668 a_{1} a_{3} + 33.234 a_{2} a_{3} + 
 16.475 a_{3}^2 \le 1$
\end{itemize}
We do not display the longer $N=4$ expression.
These strong unitarity constraints will be used to fit the experimental and lattice results
 in the BGL parametrization.

We have also considered adding the $\Lambda_b \Lambda_c$ channel to the unitarity bounds, even though heavy quark
symmetry does not relate its form factors to those of $BD$ in a direct way. Indeed
a lattice QCD calculation  of the $\Lambda_b\to \Lambda_c\ell\nu$
form factors has recently appeared \cite{Detmold:2015aaa}. Unfortunately, their precision is
not yet sufficient to evaluate their contribution to  the unitarity sum in a useful way.  

The derivation of the strong unitarity bounds makes essential use of the 
 NLO HQET relations between the
form factors, but we have seen that subsubleading ${\cal O}(1/m_c^2)$ effects can be sizeable. In order to estimate their effect we randomly vary the coefficients $b_{in}$ which appear in 
(\ref{sumstrong}) in such a way that the overall shift in the ratio between form factors 
with respect to the expressions given in Appendix A of \cite{CLN} is less than 8\%.
 Eq.~(\ref{sumstrong}) precludes any coefficient $b_{in}$ from being too large. As
the $b_{in}$s  in turn depend on the expansion coefficients $a_n$ of $f_+$,  the largest $b_{in}$ variations turn out to be incompatible either with the constraints on $a_n$ we have from lattice and experiment, or with Eq.~(\ref{sumstrong}), degrading significantly the quality of the constrained fit, in a way which becomes stronger for higher $N$.
Indeed,  unitarity is very effective in constraining the higher derivatives terms of the form factors.

We also observe that
%In order to choose the appropriate range of variation for the coefficients $b_{in}$,consider 
the HQET expressions for $F_i/f_+$ have a $z$-expansion characterized by a rapid apparent convergence with $O(1)$ coefficients. For instance, 
\be
\frac{F_2(z)}{f_+(z)}\approx 0.329 (1-1.14 z-2.38 z^2-3.61 z^3 +\dots),
\ee   
where all the powers of $z$ originate from NLO corrections to the leading HQET result.
Limiting ourselves to random variations which satisfy the previous requirement and 
preserve the order of magnitude (not necessarily the sign) of the $z$-expansion coefficients in $F_i/f_+$,
we have verified that $i)$ employing the HQET relations in the derivation of the strong 
bounds leads to global fits with nearly optimal $\chi^2$ ({\it i.e} only very few and small variations have smaller minimum $\chi^2$); $ii)$ 
 the fitted value of  $|V_{cb}|$ 
is very (fairly) insensitive to changes in the HQET relations in the $N=3(4)$ cases, with a marked preference for a slight increase.
We will therefore employ the strong unitarity bounds as we have derived them, without
assigning any uncertainty. As will be shown in 
the next Section, the difference between results obtained with weak and strong bounds is always minor. However,  there is so much missing in the weak bounds that 
it seems preferable to use our imperfect version of the strong bounds for actual fits.

\section{Inputs}

\subsection{Experimental data}
Most of the $B\to D\ell\nu$ analyses by the Aleph, Cleo,  BaBar and Belle
were performed assuming the CLN parameterization and their results are given 
in terms of $\rho_1^2$ and $\eta_{ew} {\cal G}(1) |V_{cb}|$, where ${\cal G}(1)=2\sqrt{r}/(1+r) f_+(0)$ is the zero-recoil form factor, see \cite{hfag} for a recent 
average. In the following we perform a fit to the differential $w$ distribution, and therefore 
use only results  provided by Belle and BaBar.

The BaBar Collaboration published their results on the $w$ distribution
% (see for example 0707.2758v1, 0709.1698v1, 0807.4978v1, 0809.0828v2, 0902.2660v1, 0904.4063v1, 1303.0571v1).  
in \cite{Aubert:2009ac}. The $w$ spectrum is divided into ten bins of width 0.06,  
and the results are expressed in terms of the average $|V_{cb}|{\cal G}(w)$ in each bin. 
Since this entails non-negligible finite bin size effects, we have re-expressed the data in terms of $\Delta \Gamma$  for each bin.
As BaBar analysis is tied up with the CLN parameterization,  only the statistical uncertainties are provided. 
BaBar claims a $3.3\%$  systematic error at small $w$, and   we have
extended this to the entire $w$ range, with 100\% correlation between different bins 
as recommended to us
\cite{rotondo}.   An important point  is that BaBar's last bin, $1.54\le w\le 1.60$, extends beyond the physical endpoint, $w_{max}\simeq 1.5905$. 
This has to be taken properly into account and has a non-negligible effect on the fit.

Concerning Belle, we use the results of a recent analysis \cite{Glattauer:2015teq} of
$B\to D\ell\nu$, which provides the $w$-spectrum with full 
statistical and systematic errors, and correlations. The $w$ range  is divided into ten bins as in BaBar analysis, but the last one stops at $w_{max}$. The $B$ and $D$ mass values given  in Table \ref{tab:masses} are those employed in Ref.~\cite{Glattauer:2015teq} 
and reflect the relative weight of charged and neutral  $B$ mesons in their sample. We assume that these values are suitable for BaBar analysis as well, and neglect their uncertainty. The experimental data points 
 employed in the fit can be seen in Fig.~1, where they are shown as 
 measurements of $f_+(z)$ after normalizing them to the fitted $|V_{cb}|$.

\subsection{Lattice QCD calculations}
While the experimental results are more precise at large recoil (small $q^2$), lattice QCD calculations of the form factors are performed close to zero-recoil. They have 
made substantial progress in the last two decades and the most recent calculations 
of $f_{+,0}$
\cite{Lattice:2015rga,Na:2015kha} have reached a high accuracy 
both at zero-recoil (0.8\% and 4\%, respectively) and beyond.
In fact, they represent the first unquenched calculations of these form 
factors performed at values of $w$ different from 1, which significantly reduces the uncertainty of the extrapolation from the $w$ region where most data are taken.

The FNAL/MILC Collaboration have presented results for both $f_+(w)$ and $f_0(w)$
in terms of synthetic data points
at $w=1.00$, 1.08 and  1.16 \cite{Lattice:2015rga}, with complete statistical and systematic 
uncertainties and correlations. 

The HPQCD Collaboration \cite{Na:2015kha} instead present their results for 
 $f_+(w)$ and $f_0(w)$  in terms of a  BCL parametrization with only 
 the closest $B_c$ pole, 
 whose coefficients are provided with uncertainty and correlations. Since their simulations
extend to a maximal value of $z=0.013$, corresponding to $w\approx 1.11$, we extract from 
their parameterization synthetic data for $f_{+,0}$ at $w=1.00$, 1.06 and  1.12.
HPQCD $f_{+,0}$ data points have a significantly larger error and are more correlated than those of \cite{Lattice:2015rga}. 
While  HPQCD individual points are in good agreement with those by FNAL/MILC, 
there is a mild tension between 
their $f_+$ slopes at zero recoil: 
from \cite{Na:2015kha} we find $df_+/dw|_{w=1}=-1.29(11)$, 
%[but if we take $(f_+(1.06)-f_+(1))/0.06=-1.23(10)$; NB si tratta di due derivazioni diverse, la prima dalla formula BCL di HPQCD derivata rispetto a $w$, la seconda \`e il rapporto incrementale, non il limite, quindi ci sono contributi di ordine superiore] 
while FNAL has $df_+/dw|_{w=1}=-1.42(4)$, %[ $(f_+(1.08)-f_+(1))/0.08=-1.316(34)$]
 showing  a marginal discrepancy that may require further consideration.
 We assume no correlation between FNAL/MILC and HPQCD results, although the two calculations  share some of the MILC gauge ensembles; indeed, HPQCD errors are dominated by systematics.
 The lattice data points employed in the fit can be seen in Fig.~1.

Because of their limited accuracy,
we do not include previous lattice \cite{Atoui:2013zza,deDivitiis:2007ptj} and light-cone sum rules \cite{Faller:2008tr} results on the form factors. We also do not include the HQE result, partially based on BPS symmetry,   ${\cal G}(1)=1.04(2)$ \cite{Uraltsev:2003ye}.

\section{Results and discussion}

\begin{table}
\begin{center}
\begin{tabular}{|c|c|c|c|c|c|}\hline
exp data & lattice data & N,par & $10^3\times|V_{cb}|$ & $\chi^2$ & $R(D)$ \\ \hline
all & all & 2,BGL & 40.62(98)  &  22.1/26 &  0.302(3) \\ \hline %ok 
all & all & 3,BGL & 40.47(97)  &  18.2/24 &  0.299(3) \\ \hline %ok 
all & all & 4,BGL & {\bf 40.49(97)}  &  19.0/22 & {\bf 0.299(3)} \\ \hline %ok 
Belle & all & 3,BGL & 40.92(1.12)  &  11.6/14 &  0.300(3) \\ \hline %ok N=2 error on RD 0.0031 N=3 error on RD 0.0032
BaBar & all & 3,BGL & 40.11(1.55)  &  12.6/14 &  0.301(4) \\ \hline %ok N=2 error on RD 0.0033 N=3 error on RD 0.0037
all & FNAL & 3,BGL & 40.17(1.05)  &  10.4/18 &  0.293(4) \\ \hline %ok N=2 err on RD 0.005 N=3 0.0042
all & HPQCD & 3,BGL & 40.51$_{-1.71}^{+1.82}$  &  10.1/18 &  0.299(7) \\ \hline %ok
all & all & CLN & 40.85(95)  &  77.1/29 &  0.305(3) \\ \hline %ok
all & $f_+$ only & CLN & 40.33(99)  &  20.0/23 &  0.305(3) \\ \hline %ok
all & all & 2,BCL & 40.49(98) & 18.2/26 & 0.299(3)\\ \hline
all & all & 3,BCL & 40.48(96) & 18.2/24 & 0.299(3)\\ \hline
all & all & 4,BCL & 40.48(97) & 17.9/22 & 0.299(3)\\ \hline
\end{tabular}
\caption{ Fits using different parameterizations and inputs. See text for explanations. }
\label{tab:fit}
\end{center}
\end{table}
\subsection{Fits to both lattice and experimental data}
Table \ref{tab:fit} reports the results of several fits we have performed using different parameterizations and varying inputs. The first three fits employ the BGL parameterization 
and all the experimental and theoretical inputs considered in the previous section.
For our final result we adopt the $N=4$ BGL fit, which is the one where the functional form
of $f_{+,0}$ has the maximum flexibility. 
 In the case $N=2$ the absolute minimum of $\chi^2$  is always consistent with
   both weak and strong unitarity bounds. For $N=3,4$ the absolute minimum lies outside both the weak and strong 
 unitarity bounds. We therefore look for the constrained minimum imposing the strong unitarity bounds, which modifies slightly
 the fitted values of $|V_{cb}|$ and $R(D)$ and complicates the error analysis,  
   giving rise to asymmetric uncertainties, which we evaluate using $\Delta\chi^2=1$ (.
Weak and strong unitarity constraints lead to very similar results, with $|V_{cb}|$ just $0.05\, 10^{-3}$ higher in the second case for $N=4$.
 
In all cases the fits have good quality and there is a remarkable stability with respect to the value of $N$, which could not be achieved without the unitarity bounds.
Thanks to the unitarity bounds, 
 the error on $R(D)$ is reduced by 30\% in the case of $N=3$ and by 50\% in the case of $N=4$, while that on $|V_{cb}|$ is only slightly reduced for $N\ge 3$.
 The coefficients $a_i, b_i$ of the form factors $f_{+,0}$ obtained in these three BGL 
fits are shown in Table  \ref{coeff} with their errors. In Table \ref{corr} we provide the correlation matrix
for the case $N=2$, the only case for which it can be properly defined.
\begin{table}
\begin{center}
\begin{tabular}{|c|c|c|c|c|c|}\hline
 N  &  $a_0$ & $a_1$ &$ a_2$ &$a_3$ & $a_4$\\  \hline 
 2 &$  0.01566(11)$&$ -0.0342(31)$& $-0.090(22) $& &\\  \hline
 3 &$ 0.01565(11) $ &$-0.0353(31)$  &$-0.043(^{+21}_{-35})$% -0.0431^{+206}_{-350}$ 
 &$ 0.194(^{+19}_{-16})   %0.1940^{+186}_{-1616}
 $ & \\  \hline
 4 &  
 $0.01564(11) $&$-0.0353(30)$ &$-0.044(^{+22}_{-14})% -0.0437(^{+216}_{-140})
 $ &$ 0.111(^{+51}_{-111})$%0.11144^{+5103}_{-11149}$ 
 &$-0.20(^{+20}_{-8})$%-0.1953^{+1938}_{-814}$
 \\ \hline\hline
  N  &  $b_0$ & $b_1$ &$ b_2$ &$b_3$ &$b_4$\\ \hline
  2 &$ 0.07935(58)$ &$-0.205(14)$&$-0.23(10)$ & & \\ \hline
 3 &$ 0.07932(58)$ &$ -0.214(^{+15}_{-14})$%-0.2142^{+149}_{-141}$ 
 &$0.17(^{+10}_{-25})%0.1732^{+994}_{-2462}
 $ &$-0.958(^{+1060}_{-2})%-0.9580^{+10601}_{-16}
 $ &  \\ \hline
 4 & $0.07929(^{+97}_{-93}) $&$ -0.210(14)$%-0.2096^{+137}_{-136}$ 
 &$0.09(^{+12}_{-14})%0.0909^{+1207}_{-1414}
 $&$-0.967(^{+396}_{-11})%-0.9666^{+3962}_{-105}
 $ &$0.08(^{+76}_{-71})%0.0847^{+7648}_{-7061}
 $\\ \hline
\end{tabular}
\end{center}
\caption{Coefficients of the form factors in the BGL fits to all data.}
\label{coeff}
\end{table}
In the case $N=4$  the 1$\sigma$ uncertainties for $f_{+,0}(z)$ are
\bea
\delta f_+(z)&\simeq&  0.00854 + 0.0388 z + 0.26 z^2, \nn \\
\delta f_0(z)&\simeq&   0.0065 + 0.012 z + 1.2 z^2 .
\eea
These uncertainties are very close to those we obtain from the $N=2,3$ fits.
  \begin{table}
\begin{center}
\begin{tabular}{|c|c|c|c|c|c|}\hline
  &  $a_0$ & $a_1$ &$ a_2$& $b_1$ & $b_2$\\  \hline 
 $a_0$ &$  1$&$ 0.304$& $-0.294$& 0.212&0.161\\  \hline
 $a_1$ &$ 0.304 $ &$1$  &$-0.422$ &0.747 &0.190 \\  \hline
 $a_2$ &  $-0.294 $&$-0.422$ &$1 $ &$-0.034$ &  0.148\\ \hline
  $b_2$ &  $ 0.212$&$0.747$ &$-0.034 $ &1 & $-0.210$ \\ \hline
   $b_2$ &  $0.161 $&$0.190$ &$0.148 $ &$-0.210$ & 1 \\ \hline
\end{tabular}
\end{center}
\caption{Correlation matrix for the $N=2$ fit. Due to the constraint $f_+=f_0$ at $q^2=0$, the coefficient $b_0$ 
is given by  $4.99 a_0 + 0.32 a_1 + 0.021 a_2 - 0.065 b_1 - 0.004 b_2$. }
\label{corr}
\end{table}

We also present two fits performed with the CLN parameterization.
The first one includes all experimental and lattice data and  has a very low $p$-value, 3 $10^{-6}$. This is due to the fact that in the CLN parameterization the ratio $f_+/f_0$ is fixed to the HQET relation (\ref{f0CLN}), which 
is in striking contrast with the most precise lattice evaluations: Ref.~\cite{Lattice:2015rga}
finds  $f_+(0)/f_0(0)=0.753(3)$ at zero recoil, while (\ref{f0CLN}) implies 0.775. 
The two values differ only by 3\%, which is of the expected order of magnitude 
for higher order corrections to the CLN relation. Clearly, lattice calculations 
are getting too precise to use CLN without a proper uncertainty.
A second CLN fit (last row of Table \ref{tab:fit}) which excludes all $f_0$ lattice determinations has a good quality, comparable to that of the BGL fits.  

With the BCL parameterization we have performed fits to all experimental and lattice data for $N=2,3,4$,  with weak unitarity constraints only. The results are perfectly consistent with those obtained in the BGL parameterization and are very stable for increasing $N$. 

%is certainly related to the absence of the outer functions, whose $z$-expansion has large higher order coefficients.
  \begin{figure}[t]
 \begin{center}
  \includegraphics[width=13.5cm]{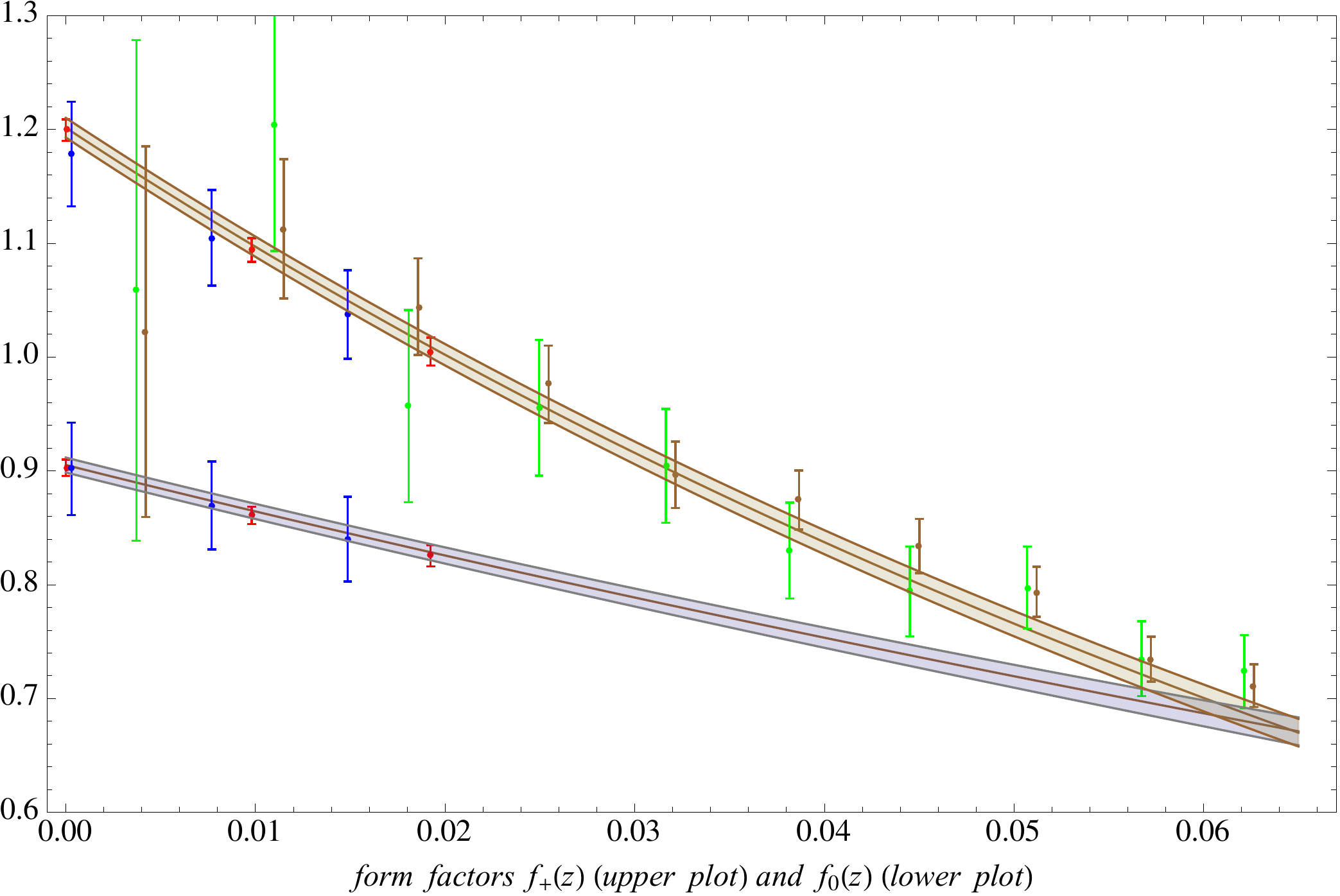} 
  \caption{Form factors in the $N=4$ fit with data points. FNAL/MILC synthetic data 
  are shown in red, HPQCD in blue, Belle data in brown, BaBar in green.}
\label{fig:fit}
\end{center}
\end{figure} 

Our fits are in good agreement with recent analyses, if one takes into account the different inputs. Belle analysis 
\cite{Glattauer:2015teq} employs the same lattice data as we do and
finds (BGL, $N=3$) $\eta_{ew}|V_{cb}|=41.10(1.14)\, 10^{-3}$. Using $\eta_{ew}=1.0066$ as 
we do, one gets $|V_{cb}|=40.83(1.13)\, 10^{-3}$, well compatible with the result of our Belle-only fit. The HPQCD collaboration reports $|V_{cb}|=40.2(1.7)(1.3)\, 10^{-3}$ \cite{Na:2015kha}  
based on Babar data only and using a different $\eta_{ew}$. If we repeat the fit under similar  conditions we get $|V_{cb}|=40.1(2.2)\, 10^{-3}$.
The FNAL-MILC collaboration quotes $|V_{cb}|=39.6(1.7)(0.2)\, 10^{-3}$ \cite{Lattice:2015rga},  based on BaBar data only and using $\eta_{ew}=1.011(5)$. Under the same hypotheses we get $|V_{cb}|=39.7(1.7)\, 10^{-3}$. A fit with lattice, BaBar, and preliminary Belle data, presented in \cite{DeTar},  is again consistent with our results after taking into account the different inputs.

It might be interesting to compare our results with those one obtains using the HFAG
average $\eta_{ew} {\cal G}(1) |V_{cb}|=42.65(72)(1.35)\ 10^{-3}$ \cite{hfag} and the 
$N=4$  fit value of $f_+(0)$, corresponding to  ${\cal G}(1)=1.0557(78)$. We get 
 $|V_{cb}|=40.13(1.47)\, 10^{-3}$, which is consistent with but less precise than our final value. This is clearly not surprising because we include new additional information. 
 One should keep in mind that the data averaged by HFAG are the result of CLN extrapolation. 

To gauge the impact of non-zero recoil lattice results in the analysis, we perform a  fit
without all the lattice points at $z\neq 0$: the result is   
%$|V_{cb}|=39.6(2.1)\, 10^{-3}$ using BGL with $N=2$, 
$|V_{cb}|=39.6(^{1.7}_{2.0})\, 10^{-3}$ using BGL with $N=3$.
%$|V_{cb}|=39.6(1.5)\, 10^{-3}$ with $N=4$. 
Clearly,  non-zero recoil lattice data are very important both for the uncertainty and the central value.
If we instead employ CLN we obtain $|V_{cb}|=40.0(1.1)\, 10^{-3}$.

The form factors of the $N=4$ BGL fit 
are shown in Fig.~\ref{fig:fit} together with their 1$\sigma$ error bands and the lattice input
data. We also show bin-average values for $f_+$ obtained from the experimental data, with normalization  fixed by the fitted value of  $|V_{cb}|$.
\vspace{4mm}

\subsection{Fits to lattice results only}
 \begin{figure}[t]
 \begin{center}
  \includegraphics[width=8cm]{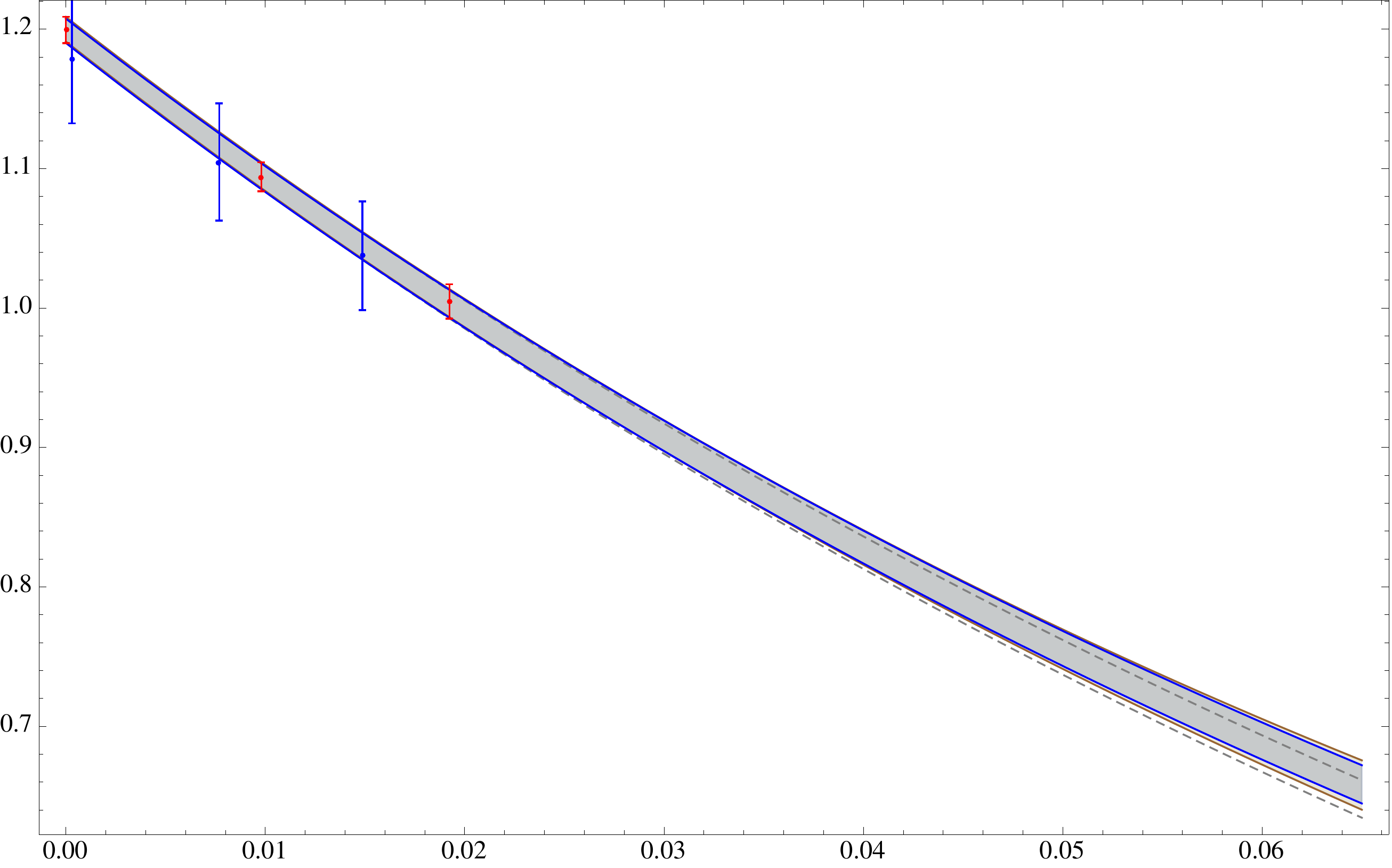}
 \includegraphics[width=8cm]{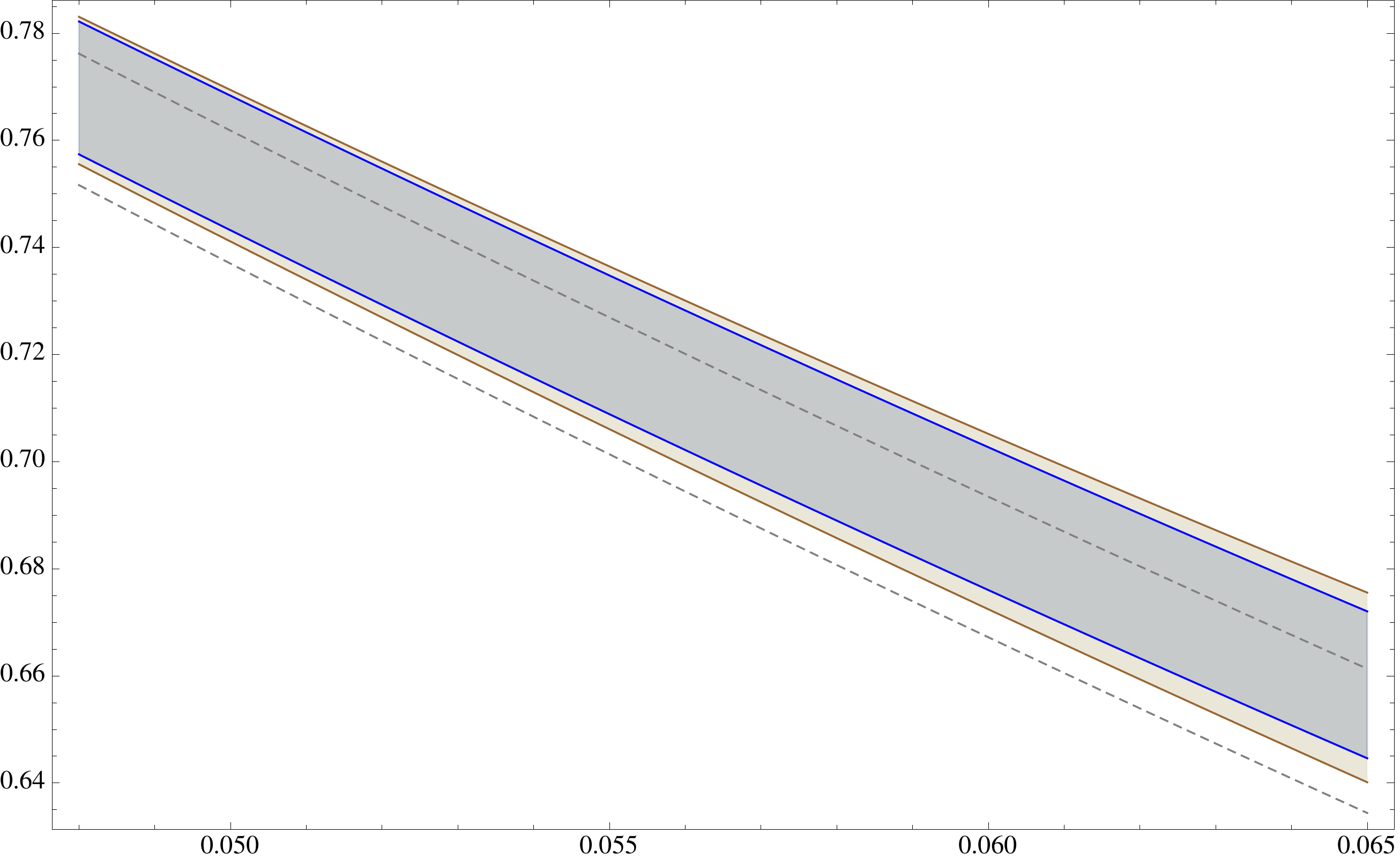}
  \caption{ Form factor $f_+(z)$ in the $N=4$ BGL fit to lattice data for $f_{+,0}(z)$ with weak (brown band) and strong (gray band) unitarity constraints. The $N=2$ band (independent of unitarity constraints) is shown in dashed lines for comparison. FNAL/MILC synthetic data 
  are shown in red, HPQCD in blue. On the right, enlarged detail of the tail. }
\label{fig:fit2}
\end{center}
\end{figure} 
We have seen that the difference between using weak and strong unitarity constraints is relatively small in our fits. One might think this is going to change in a case where we do not have data over the whole spectrum and where extrapolation errors become important. To illustrate such a case, we have performed a fit  to
lattice results only, shown in Fig.~2. Here we include the same  $f_+$ and $f_0$ results by the MILC-FNAL and HPQCD collaborations we have employed in the global fits. The plot
shows the $1\sigma $ error band for $f_+$ in the case of the $N=4$ BGL fit, using 
strong and weak constraints. The band obtained using strong constraints is up to 25\% narrower
than the one obtained using the weak constraints. These bands can be compared with similar ones given in Refs.~\cite{Lattice:2015rga,Na:2015kha}, keeping in mind that ours is a combined fit. 
%The $N=4$ band obtained without unitarity constraints covers all the region without lattice points.
We also note in passing that the implementation of weak unitarity constraints using gaussian priors, as done for instance in \cite{Lattice:2015rga}, leads to overestimate 
the width of the band (by up to 30\%  in the $N=4$ case with only  Fermilab results).
In conclusion, while for $N>2$ it is essential to use unitarity constraints, the gain from using strong rather than weak constraints is not  significant in our fits. However, this is not a general statement and the issue should be reconsidered case by  case.

\section{Summary}
We have re-examined the form factor parameterizations for $B\to D \ell\nu$ in view of recent theoretical and experimental results. After updating the unitarity 
constraints to ${\cal O}(\as^2)$ with recent quark mass values, we
 have discussed the strong unitarity bounds, which can  
 improve the precision of the form factors and lead to better determinations of 
$|V_{cb}|$ and $R(D)$. In the future, it might be possible to implement strong unitarity bounds using lattice calculations of different form factors, rather than HQET approximations only, and to perform global fits to experimental and lattice data for different channels
({\it e.g.}\ $B\to D,B\to D^*$ etc. and even $\Lambda_b\to \Lambda_c$).

 We have considered the BGL, CLN, and BCL parameterizations; they 
 all yield consistent results. However   
 the CLN parameterization, which has played a useful role in the past,  may no longer be adequate to cope with the present accuracy of lattice calculations. BGL and BCL are valid alternatives. In both cases, the fits should be performed with increasing $N$,  properly including the unitarity constraints.  
Our final result for $|V_{cb}|$,
\be
|V_{cb}|=40.49(97)\, 10^{-3},
\label{finalVcb}
\ee
has a 2.4\% error and can be improved by more precise lattice calculations  and 
by new measurements of the differential decay rate. 
It is fair to stress that the level of precision in (\ref{finalVcb}) is mostly 
due to the high precision FNAL-MILC results, which makes also urgent to have alternative 
calculations at the same level of accuracy.
Concerning experiment, even before Belle-II data are available,
 the old BaBar data could be usefully  re-examined using the latest tagging techniques and
untying them from the CLN formulas. Our result is compatible with, but less precise than 
both the exclusive $V_{cb}$ from $B\to D^*\ell\nu$ in Eq.~(\ref{excl}) and the inclusive one of Eq.~(\ref{incl}). While the  $V_{cb}$ conundrum persists, a new player has entered the game.

The HFAG average of the BaBar and  Belle measurements of $R(D)$ is \cite{hfag}
\be
R(D)_{exp}= 0.397 \pm 0.040\pm 0.028\nn
\ee
which differs from our central value 0.299(3) by 2$\sigma$. Our SM determination of $R(D)$ is the most precise so far and is in excellent agreement  
with other recent estimates: the HPQCD collaboration \cite{Na:2015kha}, 
without recourse to experimental data,
%using BaBar data, 
reports 0.300(8), while 
  FNAL-MILC \cite{Lattice:2015rga}, using BaBar data only, finds 0.299(11). Older analyses such as those of Refs.~\cite{Kamenik:2008tj,Becirevic:2012jf} give consistent values.

\subsection*{Acknowledgements}
We thank C.~Bouchard, I.~Caprini, C.~De Tar, R.~Glattauer, K.~Healey, R.~Lebed, A.~Kronfeld, A.~Polosa, M.~Rotondo,  C.~Schwanda, R.~Van de Water, and H.~Zechlin for useful discussions. We are grateful to the Mainz Institute for Theoretical Physics (MITP) for hospitality and partial support during the workshop {\it Challenges in semileptonic B decays}, which prompted us to start this work.  This research was partly conducted  at KITP and was supported in part by the National Science Foundation under Grant No. NSF PHY11-25915.

\end{document}